\documentclass[aps,prl,twocolumn,showpacs,groupedaddress]{revtex4}
\usepackage{graphicx}
\usepackage{amssymb}
\usepackage{amsmath}
\usepackage{array}
\begin{document}
\title{Pseudo-random binary injection of levitons for electron quantum optics}
\author{D. C. Glattli  and P. Roulleau}

\affiliation{\vspace{5pt} Service de Physique de l\'{ }\'{E}tat Condens\'{e} (CNRS UMR 3680), IRAMIS, CEA-Saclay,
 F-91191 Gif-Sur-Yvette, France.}

\date{\today}
\begin{abstract}

The recent realization of single electron sources lets envision performing electron quantum optics experiments, where electrons can be viewed as flying qubits propagating in a ballistic conductor. To date, all electron sources operate
in a  periodic electron injection mode leading to energy spectrum singularities in various physical observables which sometime hide the bare nature of physical effects. To go beyond, we propose a spread-spectrum approach where electron flying qubits are injected in a non-periodic manner following a pseudo-random binary bit pattern. Extending the Floquet scattering theory approach from periodic to spread-spectrum drive, the shot noise of pseudo-random binary sequences of single electron injection can be calculated for leviton and non-leviton sources. Our new approach allows to disentangle the physics of the manipulated excitations from that of the injection protocol. In particular, the spread spectrum approach is shown to provide a better knowledge of electronic Hong Ou Mandel correlations and to clarify the nature of the pulse train coherence and the role of the dynamical orthogonality catastrophe for non-integer charge injection.

\end{abstract}

 \maketitle

\section{\label{sec:level1}I. INTRODUCTION}
 The goal of Electron quantum optics is to perform with electrons quantum operations similar to those done with photons in quantum optics. Here we consider 
ballistic quantum conductors where electrons can propagate with no backscattering along electronic quantum modes in a way similar to photons propagating along electromagnetic modes. Many tools commonly encountered in optics are already available in ballistic quantum conductors. The analog of photon beam-splitters are obtained with electron beam-splitters using Quantum Point Contacts (QPC) which form a local artificial scatterer partitioning a single electron propagating on single quantum channel into transmitted and reflected channels. Combining two QPC beam-splitters in series 
provide electron analogs of optical Mach-Zehnder \cite{Ji03,Roulleau08} and Fabry-P\'{e}rot interferometers. Such intereferometers have been useful to evidence electronic interference and to quantify the degree of coherence of electronic wave-packets in the quantum conductor. To go further in the electron/photon analogy, a full Electron
Quantum Optics requires the analog of a single photon source. One appealing perspective is that the time control of single electrons let envisage their use as flying qubits \cite{flying_quBits1,flying_quBits2} where the information is encoded in the presence or absence of an electron in a quantum channel or encoded in the spin of the itinerant electron (to mimic photonic flying qubits encoded in the photon polarisation). 
Several approaches have been used for realizing the on-demand coherent injection of single electrons \cite{Herm11,McNe11,Flet13,Ryu16,Ubbe14,Feve07,Dubo13}.
Here, we consider the voltage pulse source \cite{Dubo13,Jull14} which is simpler to built and operate. It is based on voltage pulses applied on a contact to inject a single charge in the ballistic conductor. It was theoretically and experimentally shown  that for 
voltage pulses having a Lorentzian time variation, electrons are injected in the form of a remarkable \textit{minimal excitation state } \cite{coherentstates,Ivan97,lebePRB05,Keeling06,Hassler08} which has been called a \textit{leviton} \cite{Dubo13}. Synchronizing the injection of single electrons from different sources \cite{Janine09,Moskalets11} and letting them interfere in a quantum conductor lets envisage flying qubit operation in a simple way. This approach has already lead to new quantum experiments where single electron partitioning \cite{Dubo13,Parm12}, electronic Hong Ou Mandel interference \cite{Dubo13,Bocq13}, or single electron quantum tomography \cite{Jull14} have been shown. 

For practical ease of operation and calculation, only periodic electron injection have been considered to date. However periodic driving is not mandatory. What is usually needed is  a large number of single electron injections to perform with high enough accuracy the statistical measurements giving the average value of the current or of its fluctuations (current noise) when performing Hong Ou Mandel correlations or  Quantum State Tomography. Periodic driving leads to peculiar dependence to the observables studied. For example, the electron injection at frequency $\nu=1/T$
 introduces stepwise discontinuities in the electron energy distribution of levitons, also imprinted in their electronic Wigner function \cite{Grenier11,Jull14,Ferra13}, at energies multiple of $h\nu$ \cite{Dubo12}. These discontinuities manifest by singularities in the shot noise of electrons partitioned by a QPC. They may prevent to understand if some observed phenomenon results from periodicity or from the nature of the injected charge. This is unfortunate as understanding the nature the single electron state when many electrons are injected is a key issue and a theoretical challenge \cite{Moska15,Gaury14}. The ideal situation would be to inject just a single electron and look at the result. This is however impossible due to the present lack of reliable single electron detectors. 
 
 We consider here the non-periodic injection of electrons following a pseudo-random binary bit pattern $\{b_{k}\}$  where at each time $t=kT$, $k$ integer, one (or no) electron is injected if $b_{k}=1$ (or $0$), see figure \ref{PofELevit}(a). This provides a situation intermediate between the periodic and the single electron injection. Also a binary injection is what we have to be prepared to do for flying qubit operation in electron quantum optics. This is thus a field of investigation that it may be worth to develop. Regarding pure Physics concerns, a direct result of random injection is to spread the energy spectrum and remove the $h\nu$ singularities. We will show that the  electron energy spectrum is made from a continuous part directly related to the statistics of the bit ensemble $\{b_{k}\}$ while weaker spectral discontinuities remains which are directly related to the regular bit sequence. Looking at the off-diagonal element of the energy density matrix we will show that this spectral peaks are related to long range phase coherence. Comparing periodic and random injection provides a knob to better disentangle the physics of injected electrons from the injection physics. In the present work we will show that this enable us to better characterize the dynamical orthogonality catastrophe predicted in \cite{coherentstates} for non integer charge injection and to quantify the amount of electron-hole pairs created when injecting electrons with non-Lorentzian pulses. Another advantage of non-periodic injection is found in the case of electronic Hong Ou Mandel (HOM) interferometry which gives the time correlation function of the electron wavefunction  $ |\langle\psi(\tau)|\psi(0)\rangle|^{2}$. As we will later show, this allows for a better exploration of the tails of the single electron (leviton) wavefunction while periodicity limits this information to a half period time-scale: $\tau\leq T/2$. 
 
 In a broader perspective, non-periodic injection belongs to the class of spread-spectrum techniques used in communications and the problem provides a bridge between quantum physics and telecommunication problems. Indeed some of the theoretical results derived here are directly imported and adapted from the field of digital communications \cite{Proakis}. Considering the rapid development of periodically or quasi-periodically driven Hamiltonian \cite{Dahlhaus11,MartinHalperin17}, the study of the response of quantum systems to non-periodic spread spectrum excitations is only starting \cite{Safi14}. It may be viewed as a new field which, except in quantum communication, has not yet been explored and which could shed a new (white) light on quantum effects.

\section{\label{sec:level2}II. BACKGROUND}
\subsection{\label{sec:citeref}A. Floquet scattering approach}
Before presenting our approach to deal with this new situation, we recall the main results of the Floquet scattering approach \cite{Moska02} for periodic excitation. The next section will extend the Floquet approach to non-periodic drive. 

\begin{figure}
  \includegraphics[width=8.5cm,keepaspectratio=true,clip=true]{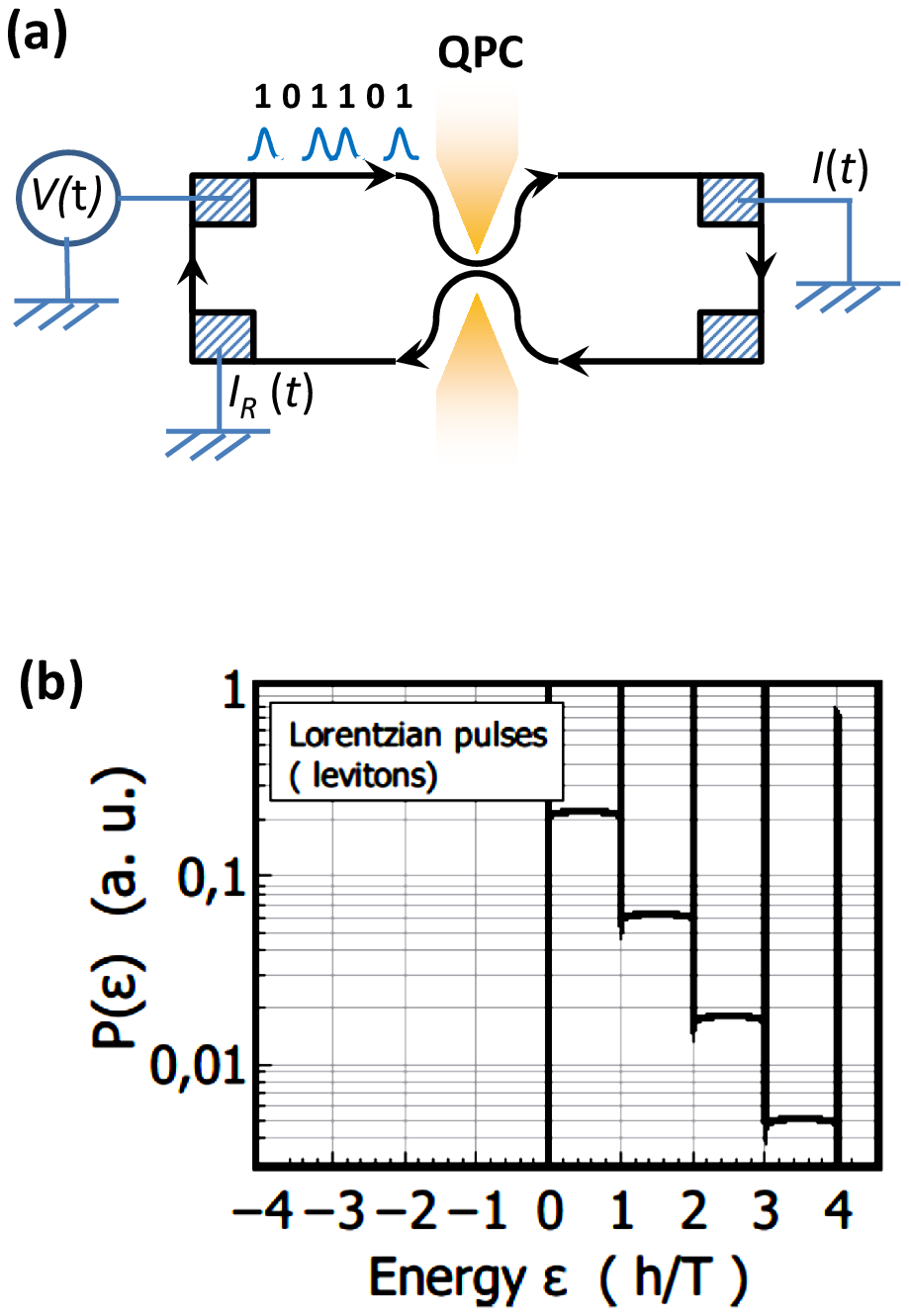}
  \caption{(a) chiral ballistic conductor with a Quantum Point Contact in its middle playing the role of a beam-plitter of transmission $D$. The voltage $V(t)$ injects levitons by a series of Lorentzian voltage pulses following a pseudo-random binary bit sequence.(b) The probability $P(\varepsilon )$ to shift by $\varepsilon $ the energy of electrons experiencing a pseudo random binary sequence of Lorentzian voltage pulses is computed from the F.T. of Eq. (\ref{Ctttimeaverage}). The single-side band energy spectrum ($P(\varepsilon )=0$ for $\varepsilon <0$) is characteristic of Lorentzian voltage pulses. It warranties that only electron like excitations are generated by the pulses as expected for leviton generation. This generalizes to random emission results found for periodic injection. $W=0.1 T$ here.}
  \label{PofELevit}
\end{figure}

We consider a periodic voltage $V_{L}(t)$ applied to, say the left, contact of a ballistic conductor while the opposite (right) contact is kept to ground ($V_{R}=0$). For simplicity spin is disregarded and the conductor is made of a single quantum channel (for example the edge channel of the integer quantum Hall effect as sketched in Fig.\ref{PofELevit}(a)). Later including the spin and generalizing to several electronic modes are straightforward. 

According to the scattering approach developed by Moskalets and B\"{u}ttiker \cite{Moska02}, an electron emitted at energy $E$ below the Fermi energy $\mu_{L}$ from the left contact and experiencing the ac excitation  has its phase $\phi(t)$ modulated by the voltage, with 
\begin{equation}\label{phioft}
    \phi(t)=\int_{-\infty}^{t} eV(t')dt'/h
\end{equation}
The time dependence breaking energy conservation, the emitted electron is scattered into a superposition of quantum states at energies $E + l h\nu$, where $l$ is an integer.
 The scattering amplitudes connecting initial and final energies are the photo-absorption (emission) amplitudes $p_{l}$ which form the elements of the so-called unitary Floquet scattering matrix, where positive (negative) $l$ means an electron being absorbing (emitting) $l$ photons. The $p_{l}$ are given by the Fourier transform of the phase term: 
 \begin{equation}\label{pl}
    p_{l}=\frac{1}{T}\int_{0}^{T}\exp(-i\phi(t))\exp(i2\pi l \nu t)dt
\end{equation} 
From unitarity, the $p_{l}$ obey the following useful identity $\sum_{l}p_{l+k}^{\ast}p_{l}=\delta_{k,0}$,where $^{\ast}$ denotes the complex conjugate. To calculate transport properties, the standard annihilation fermionic operators $\hat{a}_{L}(E )$ which operate on the equilibrium states of the left contact are replaced by new annihilation fermion operators
$\hat{\tilde{a}}_{L}(E )$, with $\hat{\tilde{a}}_{L}(E )=\sum_{l}p_{l} \hat{a}_{L}(E - lh\nu)$. This substitution on transport formula provides direct expression for the mean photo-assisted current $\tilde{I}$ and mean photo-assisted shot noise $\tilde{S}_{I}$ related to the dc transport expressions $I$ and $S_{I}$.  Defining $P_{l}=|p_{l}|^{2}$ the probability to absorb or emit $l$ photons and, anticipating later the case of non-periodic drive, we define the density of probability per unit energy as 
 \begin{equation}\label{Pl}
    P(\varepsilon) = \sum_{l} P_{l} \delta(\varepsilon-lh\nu)
\end{equation}
We can rewrite the known expression for the average current,  the average current  noise \cite{Lesovik94,Pedersen98} and the electron energy distribution under ac excitation as follows:
\begin{equation}\label{Ipheps}
    \tilde{I}(V_{dc})= \int_{-\infty}^{+\infty} P(\varepsilon ) I(V_{dc}+\varepsilon )d\varepsilon
\end{equation}
\begin{equation}\label{SIpheps}
    \tilde{S_{I}}(V_{dc})= \int_{-\infty}^{+\infty} P(\varepsilon ) S_{I}(V_{dc}+\varepsilon )d\varepsilon
\end{equation}
\begin{equation}\label{ftildeeps}
   \tilde{f}(\varepsilon )=\int_{-\infty}^{+\infty} P(\varepsilon ') f(\varepsilon-\varepsilon '-eV_{dc})d \varepsilon '
\end{equation}
where $f(\varepsilon)$ is the equilibrium Fermi distribution with energies referred to the right contact Fermi energy and  we have added an extra dc voltage $V_{dc}$ to the ac voltage $V(t)$ for generality. A typical application is a single channel conductor with a QPC in its middle transmitting electrons with transmission probability $D$. For energy independent transmission, giving linear I-V characteristics, and for $V(t)$ having zero mean value, remarkably $ \tilde{I}(V_{dc})=I(V_{dc})$ where $I(V_{dc})=D(e^{2}/h)V_{dc}$ is, according to the Landauer formula, the dc current that we would measure when only a dc bias $V_{dc}$ was applied on the left contact. Indeed from unitarity $\sum_{l} P_{l}=1$, while $\sum_{l} l P_{l}=0$. By contrast, the shot noise shows a non-linear (rectification like) variation with current (or voltage). Indeed, the zero temperature shot noise is $S_{I}=2e\frac{e^{2}}{h}|V_{dc}|D(1-D)$ \cite{MartPRB92,ButtPRB92,Blant00} where the $D(1-D)$ factor encodes the partitioning of electron between transmitted and reflected states with binomial statistics. From Eq.(\ref{SIpheps}), the singular $P(\varepsilon)$ for periodic drive, gives replica of the zero bias singularity  in the photo-assisted shot noise $\widetilde{S}_{I}$ each time $V_{dc}=lh\nu$ \cite{Lesovik94,Pedersen98,Rychkov05,Vanevic07,Bagrets07}. The singularities result from stepwise variations of the energy distribution $\tilde{f}(\varepsilon )$ of the periodically driven Fermi sea given by:
\begin{equation} \label{ftilde}
\tilde{f}(\varepsilon ) = \sum_{l}P_{l} f(\varepsilon -lh\nu - eV_{dc})
\end{equation}
where $f(\varepsilon )=1/(1+\exp(\beta(\varepsilon-\mu))$ is the Fermi Dirac distribution of electrons at electronic temperature $k_{B}T_{e}=1/\beta$ $\mu$, the right contact Fermi energy taken as energy reference. Terms with positive (negative) $l$ describe electron (hole) like excitations.

As shown in \cite{coherentstates,Ivan97,Keeling06} and detailed in \cite{Dubo12}, a particular situation arises when the voltage V(t) is a sum of periodic Lorentzian pulses 
\begin{equation} \label{Vperiod}
V(t)=\frac{h}{\pi e T}\sum_{k=-\infty}^{+\infty} \frac{1}{1+(t-kT)^{2}/W^{2}}
\end{equation}
each introducing a single electron in the conductor.
Here $P_{l}$ remarkably vanish for $l<0$ and $\tilde{f}(\varepsilon)$ describes a pure electron excitation population with no holes. This defines a \textit{minimal excitation states} \cite{Keeling06} made of a periodic train of \textit{levitons} \cite{Dubo13}. When partitioning the leviton train, the shot noise being proportional to the total number of excitations \cite{Dubo12} is minimal due to the absence of holes. In general, arbitrary shape (non Lorentzian) voltage pulses give non-zero $P_{l}$ for negative $l$ and thus generate a mixture of electron and hole excitations. This is for example the case for  sine-wave voltage pulses $V(t)=(h\nu/e)(1-\cos(2\pi\nu t))$ injecting single electrons. The qualitative difference between Lorentzian voltage pulses and other pulse shapes is that the modulated phase gives a Single Side Band (SSB) energy spectrum (only positive (negative) energies for electron(hole)-like levitons). All other kind of modulation gives a double side band spectrum. To have SSB spectrum property with positive (negative) energy requires that $\exp(i\phi(t))$ has no poles in the upper (lower) half complex plane. This is the case for the periodic injection of levitons where
\begin{equation}
\label{periodicphase}
\exp(i\phi(t))=\frac{\sin(\pi(t+iW)/T)}{\sin(\pi(t-iW)/T)}
\end{equation}
shows periodically spaced poles in the upper complex plane. 

\subsection{\label{sec:citeref2A}B. Non-periodic Floquet scattering}
We now address non periodic excitations leading to spread spectrum property. As the term 'Floquet' is associated with purely periodic phenomena, 'non-periodic Floquet scattering' may appear as an oxymoron. However the Floquet concept has been already extended beyond periodicity. Recently there has been a considerable activity regarding topological phase transitions where, using incommensurate frequencies, the bi-harmonic driving of Hamiltonian leads to new topological Floquet lattices and energy bands \cite{Dahlhaus11,MartinHalperin17}. In fact, for quantum systems the Floquet approach is essentially a way to describe scattering in energy. This is why, for quantum conductors, the Moskalets B\"{u}ttiker Floquet scattering approach \cite{Moska02} integrates so well within the Landauer-B\"{u}ttiker scattering transport approach. 

For non periodic driving, the photo-absorption probabilities become a continuous function of the energy. In this spread spectrum situation, the definition of the  photo absorption or emission probability amplitudes become 
\begin{equation}\label{pofeps}
    p(\varepsilon )=\int_{-\infty}^{+\infty}dt \exp(-i\phi(t))\exp(i\varepsilon t/\hbar)
\end{equation}
This describes the amplitude to find an electron, initially emitted by the reservoir at energy $E$, scattered to the energy $E+\varepsilon$. From this one can calculate the statistical average of $|p(\varepsilon)|^2$ giving the probability $P(\varepsilon)$ from which current, shot noise and energy distribution can be calculated using expressions similar to Eqs. (\ref{Ipheps})(\ref{SIpheps})and (\ref{ftildeeps}). A similar use of these expressions for fully random excitations and appropriate for interacting systems can be also found in \cite{Safi14} .

\section{\label{sec:level1}III. BINARY INJECTION OF LEVITONS}
\subsection{\label{sec:level3A}A. Energy scattering probability}
The pseudo-random injection leads to a continuous spectrum characterized by the energy scattering probability $P(E)$ that we derive in this section and which is the analog of the photo-absorption/emission probability considered above for periodic voltage drive. 

As a realistic practical example, we consider the injection of single charge Levitons following a pseudo-random sequence of binary bits $b_{k}=0,1$. The voltage drive applied on the left contact is thus: 
\begin{equation}\label{RndLor}
    V(t)=\frac{h}{\pi e T}\sum_{k=-\infty}^{+\infty} \frac{b_{k}}{1+(t-kT)^{2}/W^{2}}
\end{equation}
The mean value of the drive is $\langle b_{k} \rangle eh/T$ where $\langle b_{k} \rangle$ is the ensemble average of the bit value, typically 0.5 for equal bit pseudo-random probability. The phase term $\exp(i\phi(t))$ corresponding to the pseudo-random binary levitonic drive is given by:
\begin{equation}
\label{signalBinary}
\exp(i\phi(t))=\prod_{k=1}^{N}\bigg(\frac{t-kT+iW}{t-kT-iW}\bigg)^{b_{k}}.
\end{equation}
 The problem of calculating $P(\varepsilon )$ is similar to the calculation of the power spectrum density in digital communication for binary phase pulse modulation, see \cite{Proakis,Fares17}. In this analogy, the electronic quantum phase corresponds to the modulated phase,the voltage pulses to frequency pulses, the energy to frequency and $P(\varepsilon)$ to  electromagnetic power. To obtain $P(\varepsilon)$, one needs the two-time autocorrelation function:
\begin{equation}\label{Ctt'}
    \langle C(t,t')\rangle=\langle\exp(-i\phi(t))\exp(i\phi(t'))\rangle
\end{equation}
where $\langle.\rangle$ means ensemble averaging over statistically independent bit patterns. One can show that $\langle C(t,t')\rangle $ is cyclo-stationary, i.e. $\langle C(\bar{t}+\tau/2,\bar{t}-\tau/2)\rangle $ is invariant when $\bar{t}\rightarrow \bar{t}+T$, where $\bar{t}=(t+t')/2$. is the mean time and $\tau=t-t'$ the time difference. After averaging over $\bar{t}$ we get the time average correlation function $\langle\overline{C(\tau)}\rangle$ whose Fourier Transform (F.T.) gives
\begin{equation}\label{Pesprandom}
    P(\varepsilon )=(h/T)\int_{-\infty}^{\infty}\langle \overline{C(\tau)}\rangle e^{i\varepsilon \tau/\hbar}d\tau
\end{equation}

We now derive the expressions from which these physical quantities can be calculated. We will set $T=1$ and  $\tau$ and $\bar{t}$ in period units and $w=W/T$ and we consider the ensemble $\{ {b_{k}} \}$ as uncorrelated Bernouilli random variables with probability $P[b_{k}=1]=p$ and  $P[b_{k}=0]=1-p$. 

The two-time autocorrelation function is:
\begin{equation}
\label{2timeC}
C(t-\frac{\tau}{2},t+\frac{\tau}{2})=\prod_{k=-\infty}^{\infty}\bigg( \frac{t_{k}-\frac{\tau}{2}-iw}{t_{k}-\frac{\tau}{2}+iw}
.\frac{t_{k}+\frac{\tau}{2}+iw}{t_{k}+\frac{\tau}{2}-iw}\bigg)^{b_{k}},
\end{equation}
where $t_{k}=t-k$. After averaging over statistically equivalent bit sequences, we get 
\begin{eqnarray}
\label{Exper2timeC}
\langle C(\bar{t}+\tau/2,\bar{t}-\tau/2)\rangle & = & \prod_{k}^{}\bigg(1-p+ p
\frac{t_{k}^2-(iw+\frac{\tau}{2})^2}{t_{k}^2-(iw-\frac{\tau}{2})^2}\bigg)\\
& = & \frac{\sin(\pi (t-\theta_{p}(\tau)) \sin(\pi (t+\theta_{p}(\tau))}{\sin(\pi(t-iw+\frac{\tau}{2}) \sin(\pi(t+iw-\frac{\tau}{2})}\nonumber
\end{eqnarray}
 where $\theta_{p}(\tau) = \sqrt[]{\frac{\tau^{2}}{4} - w^{2}-(1-2p)iw\tau}$.
 Note that Eq.(\ref{Exper2timeC}) interpolates between no modulation ($p=0$, all $\{b_{k}=0\}$ and $C$ is constant) and the periodic case ($p=1$).
Remarkably, Lorentzian pulses (levitons) give rise to analytical expressions of the correlation functions. The two-time correlation function can be written as:
\begin{equation}\label{Cttanalytic}
    \langle C(\bar{t}+\tau/2,\bar{t}-\tau/2)\rangle = \frac{\cos(2\pi\theta_{p}(\tau))-\cos(2\pi\bar{t})}{\cos2\pi(\tau/2 - i w)-\cos(2\pi\bar{t})}
\end{equation}
As expected, it shows cyclo-stationary property.
Its time average is:
\begin{equation}\label{Ctttimeaverage}
    <\overline{C(\tau)}>= \frac{\cos(2\pi\theta_{p}(\tau))-\cos(\pi(\tau - 2i w))}{\sin (\pi(\tau - 2i w))}
\end{equation}
The first part with the square root argument in the cosine is responsible for a continuous spectrum as expected for random excitation and depends, via the parameter $p$, on the statistics of the bit injection. The second part represent harmonic contributions giving lines in the energy spectrum reflecting the regularity of injection. Finally, the imaginary part in the argument of the sine term in the denominator indicates that $C(\tau)$ has only poles in the upper half complex plane and no poles in the lower part and ensures that $P(\varepsilon )$, its F.T., vanishes for negative energies, as expected for levitons. Figure \ref{PofELevit}(b) shows the variations of $P(\varepsilon )$ for $w=W/T=0.1$ and $p=1/2$ corresponding to equal bit probability. The energy spectrum is continuous with spectral lines at energies multiple of $h/T$ and only extends to positive energies.
\begin{figure}
  \includegraphics[width=7cm,keepaspectratio,clip]{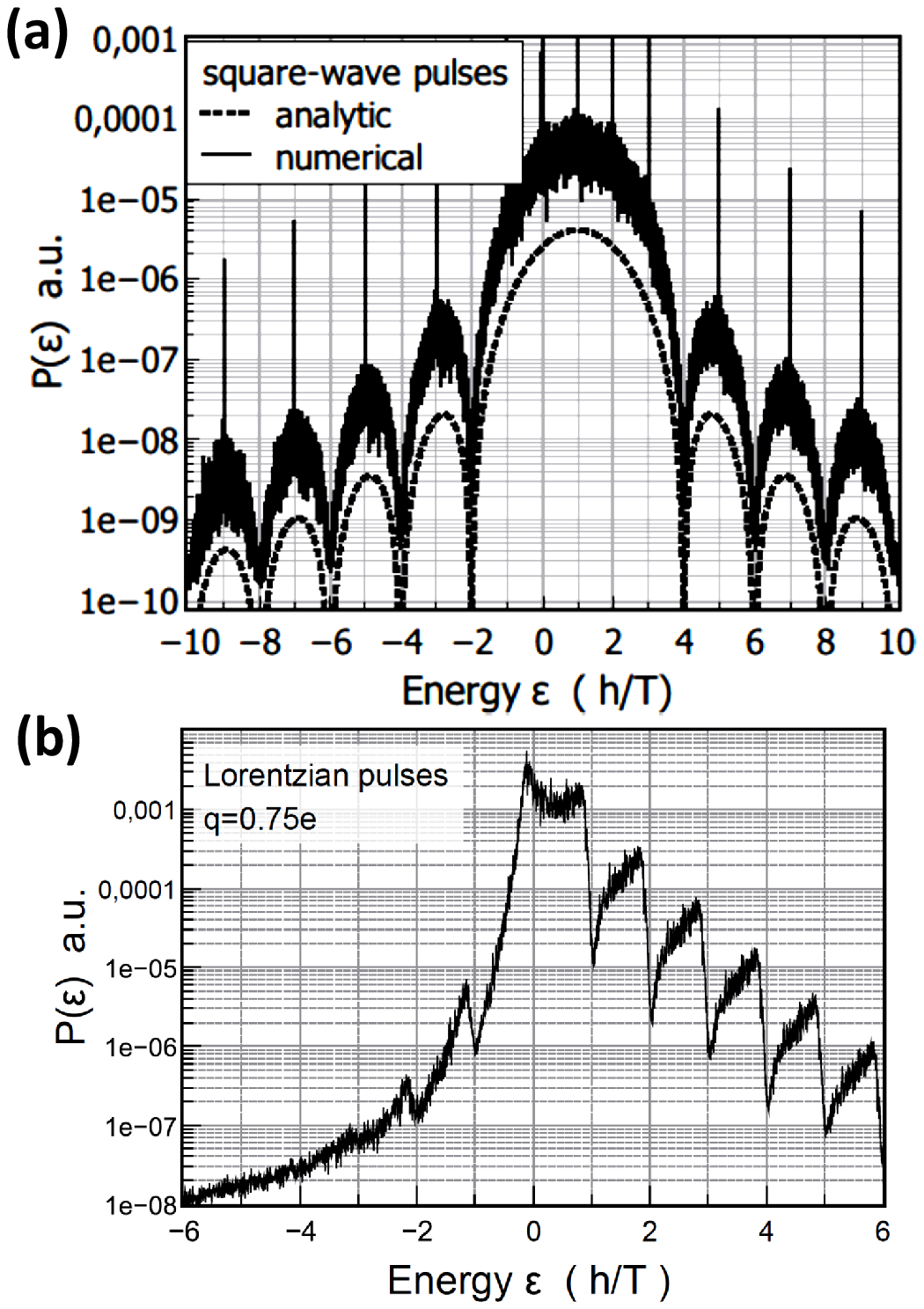}
  \caption{(a) $P(\varepsilon )$ for single electrons injected with pseudo-random binary square-wave voltage pulses of length $T/2$. The dashed curve is the continuous part of the analytic expression (\ref{NewPSquare}). The upper noisy curve is from numerical simulations, where 8 spectra made of $1024$ uncorrelated bit patterns are averaged. For clarity a vertical shift between curves has been made by applying a different arbitrary scale factor to the curves. The double-side band energy spectrum shows that square-wave pulses, unlike levitonic pulses, generate hole like excitations as already observed for periodic injection. 
  (b) $P(\varepsilon )$ for binary random Lorentzian voltage pulses injecting a non-integer charge. The absence of sharp peaks observed for integer charge is characteristic of fractional charge pulses. Although the pulses are Lorentzian non-integer pulse are not minimal excitation states as the energy spectrum is double side-band signaling hole-like excitations.}
  \label{PofESquare}
\end{figure}

For comparison we consider similar pseudo-random binary injection of single electrons but with square wave voltage pulses, with
$V(t)=b_{k}V_{0}(t-kT)$ where $V_{0}(t)=(2h/eT)$ if $t\in [0,T/2]$ and $V_{0}=0$ for $t\in [T/2,T]$. Expressing $\varepsilon $ in reduced unit of $h/T$, we found $P(\varepsilon) $ given by:
\begin{equation} \label{NewPSquare}
\begin{split}
P(\varepsilon)& =  \frac{1}{4}\bigg[\bigg(\frac{\sin(\pi \varepsilon/2)}{(\pi \varepsilon/2)(2-\varepsilon )}\bigg)^{2}+\frac{9}{4}\delta(\varepsilon )+ ... \\
 & ...+\frac{1}{4} \delta(2-\varepsilon )+ \sum_{p=-\infty}^{+\infty} \frac{4}{\pi^{2}}\frac{1}{(2p^{2}-1)^{2}} \delta(2p+1-\varepsilon ) \bigg]
\end{split}
\end{equation}
As for the levitons, the first term gives a continuous energy spectrum followed by a series of lines at multiple of the characteristic energy $h/T$. One
can show, see \cite{Proakis}, that the first term is $\propto \langle b_{k}^2\rangle - \langle b_{k}\rangle^{2}$ , $=1/4$ here, and results from the fluctuating part of the injection making the spectrum continuous, while the next terms, the spectral line, $\propto \langle b_{k} \rangle^{2}$, result from the regular part of the injection.
Figure \ref{PofESquare}(a) shows the continuous part of $P(\varepsilon )$ calculated from Eq.(\ref{NewPSquare}) and the results of numerical simulations.

The  Dirac peaks in the energy scattering probability $P(E)$ are not the sole result of regular injection but are intimately related to the integer charge. From the field of digital communication, where the bit information is coded by phase modulation, it is known that similar peaks in the emission power of digital phase modulation also appear when the so-called modulation index $h$ is integer, i.e. when the phase increment $\Delta\phi=h2\pi$ associated with an elementary frequency pulse is multiple of $2\pi$, see \cite{Proakis} chapter 3. A $2\pi$ phase increment is exactly what is imprinted to the electronic phases when injecting integer charges. Indeed using $I(t)=\frac{e^{2}}{h}V(t)$, one finds $\Delta\phi/2\pi=\int_{0}^{T}dteV(t)/h=\int_{0}^{T}I(t)dt=q$. To suppress the peaks which carry useless energy, telecommunication engineers prefer to use a non integer modulation index. In electron language, suppression of peaks will occur for non-integer charge injection as the electronic wave modulation index is $h=q/e$. This is shown in figure \ref{PofESquare}(b) for random binary Lorentzian voltage pulses carrying charge $q=0.75e$.  We see here an example where non periodic injection leads to physical manifestations not observable in the case of simple periodic injection.

\subsection{\label{sec:level3B}B. Quantum coherence of pseudo-random injection}
From a superficial inspection one may conclude that the random injection, like thermal injection, would lead to incoherent charge states. Here, we show that this is not the case. As it has been observed in the case of periodic injection, the density matrix in energy representation presents significant off-diagonal components, a signature of coherence of the injected leviton train. While for the thermal emission of electrons by a contact there is no phase relation between two injected electrons, by contrast the regular pseudo-random injection of integer charge is shown to preserve phase coherence between electrons. 

Lets first consider the case of pure periodic injection. We start from the first order electron coherence $g_{1}(t',t)=\langle \psi^{\dagger}(t')\psi (t)\rangle$ \cite{Moska15,Gren11,Haack13} where the brackets here means quantum statistical average and the Fermionic electron operator $\psi$ describing the injected electrons is $\psi (t)= \int_{-\infty}^{+\infty}d\varepsilon \hat{\tilde{a}}_{L}(E ) e^{-i\varepsilon t}$ . We consider linearized dispersion relation and $t$ is a short notation for $t-x/v_{Fermi}$ and $\hat{\tilde{a}}_{L}(E )$, defined above, describes the annihilation operator of electrons having experienced the voltage pulse. Using energy representation $\psi(\varepsilon_{0})= \int_{-\infty}^{+\infty}dt\psi(t)e^{i\varepsilon_{0}t}$, we will focus on the elements $\rho(\varepsilon',\varepsilon)$ of the electron density matrix  in energy representation:
\begin{equation}\label{densitymatrix}
    \rho(\varepsilon',\varepsilon)= \langle  \psi^{\dagger}(\varepsilon') \psi(\varepsilon)     \rangle - \langle  \psi^{\dagger}(\varepsilon') \psi(\varepsilon)     \rangle_{FS}
\end{equation}
It is directly related to the the first order coherence via $g_{1}(t',t)=\int_{-\infty}^{+\infty}d\varepsilon'd\varepsilon e^{i(\varepsilon't'-\varepsilon t)}\rho(\varepsilon',\varepsilon)$ . The right term of the right hand side represents the subtraction of the Fermi Sea (FS) contribution. Using the expression of $\hat{\tilde{a}}_{L}(E )$ for periodic injection, one finds: 
\begin{equation}\label{densitymatrix}
    \rho(\varepsilon',\varepsilon)= (\tilde{f}(\varepsilon)-f(\varepsilon))\delta(\varepsilon-\varepsilon') + \sum_{k\neq 0 } \tilde{f}_{k}(\varepsilon) \delta(\varepsilon'-\varepsilon-kh\nu) 
\end{equation}
The first term of the right hand side equation is the diagonal term representing the energy distribution as in Eq.(\ref{ftildeeps}) but with subtracted FS. The last term with $\tilde{f}_{k}(\varepsilon)=\sum_{l}p_{l-k}^{\ast} p_{l}f(\varepsilon-lh\nu)$ describes the non-diagonal terms. They characterize the coherence of the injection. They have been recently measured in order to perform the Quantum State Tomography of periodic leviton trains in Ref.\cite{Jull14}. The non-diagonal terms are responsible for long range correlation between times $t'$ and $t$ in $g_{1}(t',t)$. 

The fact that, for periodic drive, there are non zero diagonal terms $\tilde{f}_{k}$ linking energies separated by $kh\nu$ was found in deriving Eq.(\ref{densitymatrix}) originating from the cyclostationary of two-time correlation function  $C(t',t)=exp(-i\phi(t))exp(i\phi(t')$ with mean time $\bar{t}=(t+t')/2$ and relative time $\tau=t-t'$. 
\begin{equation}\label{densitymatrixtt'}
    C(t',t)=\sum_{k} C_{k}(\tau) e^{-ik2\pi\nu \bar{t}}
\end{equation}
where $C_{k}(\tau)=e^{ik\pi\nu\tau} \sum_{l}p_{l-k}^{\ast} p_{l}e^{il2\pi\nu \tau} $. 

Similarly,  for pseudo-random leviton injection we have found that the two-time correlation function $C(t',t)$ is cyclostationary after averaging over an ensemble of random bit realizations. From this observation, we can define the following relative time correlation functions $\langle C_{k}(\tau) \rangle$ :
\begin{equation}\label{Ctt'}
    \langle C(t',t) \rangle =\sum_{k}  \langle C_{k}(\tau) \rangle  e^{-ik2\pi\nu \bar{t}}
\end{equation}
where we found that $\langle C_{k}(\tau) \rangle=e^{ik2\pi\tau}e^{-k2\pi w} \langle C_{0}(\tau) \rangle$ and $\langle C_{0}(\tau) \rangle$ has been given in Eq.(\ref{Ctttimeaverage}):
We show here that despite the injection randomness the cyclostationary property leads to finite off diagonal density matrix elements for energy separated by the quantity $kh\nu$. 

The continuous version of Eq.(\ref{densitymatrix}) is given by 
\begin{equation}\label{continuousdensitymatrix}
    \rho(\varepsilon',\varepsilon)= \int_{-\infty}^{\infty}dE \tilde{f}_{E}(\varepsilon) \delta(\varepsilon'-\varepsilon-E)
\end{equation}
where the non-diagonal energy density term is:
\begin{equation}\label{continuousdensitynondiag}
    \tilde{f}_{E}(\varepsilon)=\int_{-\infty}^{+\infty}dE' p^{\ast}(E'-E) p(E')f(\varepsilon-E')
   \end{equation}
   and $p(E')=\frac{1}{2\pi}\int_{-\infty}^{\infty}dt e^{-i\phi(t)}e^{iE't}  $. Averaging Eq. (\ref{continuousdensitynondiag}) over bit ensembles and using (\ref{Ctt'}) we find that:
  \begin{equation}\label{contdensitynondiag}
   \tilde{f}_{E}(\varepsilon)=\frac{\delta(E-kh\nu)}{2\pi} \int_{-\infty}^{+\infty}dE' \int_{-\infty}^{+\infty}d\tau  \langle C_{k}(\tau)\rangle f(\varepsilon-E')
\end{equation}
Thus, non-diagonal elements connect energies separated by the amount $h\nu$ as in the periodic drive case.

Having shown the coherent character of a train of randomly injected levitons, we would like to discuss its meaning. Off diagonal coherence is related to cyclostationarity of the two-time correlation functions of the phase variation imprinted on the electronic wavefunction. It is related to the regular periodic bit injection. Would the time $T$ between two injections be random, the cyclostationary character of the injection statistics would be lost. The regular injection ensures that two levitons injected at distant times keep a well defined temporal phase and hence preserves some coherence property.  

Finally, the notation of $P(E)$, which describes scattering in energy, may confuse the reader with the $P(E)$ notation used to describe dynamical Coulomb blockade (DCB). In the latter case $P(E)$ describes the energy scattering due to the back-action of electron shot noise which in finite impedance external circuit transforms into random incoherent voltage fluctuations. Because of the fully random voltage fluctuations no off-diagonal density matrix elements are to be expected and the phenomenon is purely incoherent. We consider here the only useful experimental situation where the external circuit impedance, typically $~50$Ohms, is negligible and this DCB interaction regime is disregarded.

\section{\label{sec:level4}IV. QUANTIFYING ELECTRON-HOLE EXCITATIONS}
\subsection{\label{sec:level4A}A. Levitons versus non minimal integer charge pulses}
It is interesting to quantify the number of excitations per pulse generated by the random pulses and compare with the periodic case. The number of electrons (holes) $N_{e}$ ($N_{h}$) created per pulses is given by $N_{e}=\int_{0}^{\infty}\varepsilon P(\varepsilon )d\varepsilon $  ($N_{h}=\int_{-\infty}^{0}(-\varepsilon ) P(\varepsilon )d\varepsilon $) while the average charge per pulses is $\langle q\rangle =e(N_{e}-N_{h})$. With these definitions, the number of extra excitations accompanying the injected charge is $\Delta N_{exc}=N_{e}+N_{h}-\langle q\rangle$ per pulses. In terms of $P(\varepsilon )$:
\begin{equation}\label{meanQ}
    \langle q\rangle= \int_{-\infty}^{+\infty} P(\varepsilon )\varepsilon d\varepsilon
\end{equation}
\begin{equation}\label{Neh}
   N_{e}+N_{h}= \int_{-\infty}^{+\infty} P(\varepsilon )|\varepsilon | d\varepsilon
\end{equation}
 $ N_{e}+N_{h}$ can be measured using Shot Noise measurements, see Eq.(\ref{SIpheps}). This was  experimentally demonstrated in \cite{Dubo13}. Indeed, if the charges injected by the pulses in a ballistic conductor are send to a beam-splitter of transmission D, their partitioning gives the  zero temperature noise $S_{I}=2(e^{2}(e/T)D(1-D)(N_{e}+N_{h})$ \cite{Dubo12}.
 \begin{figure}
  \includegraphics[width=7cm,keepaspectratio,clip]{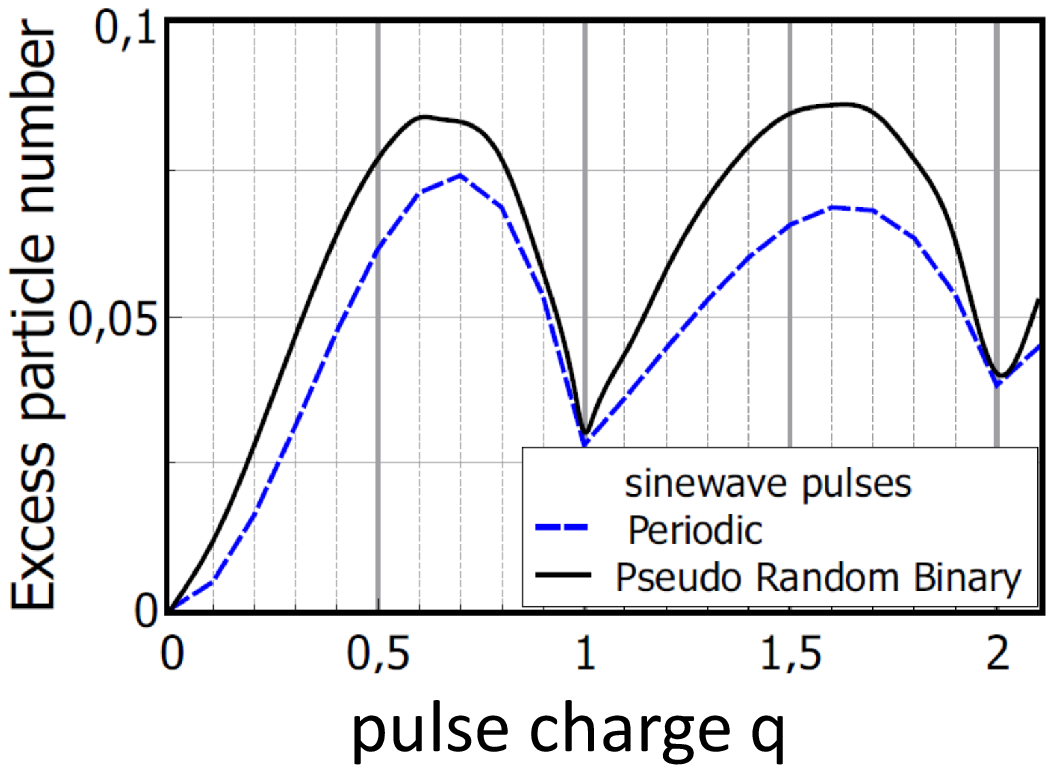}
  \caption{The figure compares the excess number of excitations $\Delta N_{exc}$ produced per period for sinewave pulses carrying charge $q$ for periodic and binary pulse injection. $\Delta N_{exc}$ shows local minima for integer charges while for non-integer charge it grows as a result of the Dynamical Orthogonality Catastrophe (DOC). Random binary injection giving more excitations than periodic injection is consistent with the DOC scenario as, in average,separation between pulses is larger.}
  \label{cata}
\end{figure}
 We now compare various single charge voltage pulses. For all pulse shapes, the computation of (\ref{meanQ}) from $P(\varepsilon )$ gives the same average charge $\langle q\rangle=e/2$ per injection period, trivially resulting from equal bit probability. The Lorentzian pulses give no extra excitation $\Delta N_{exc}=0$ as expected while the square wave pulses
 give $\Delta N_{exc}/<q>=0.1289$. This is to be compared with the periodic square wave case for which it was found in \cite{Dubo12,Dubo13}  $\Delta N_{exc}/<q>=0.1082$. We have also numerically computed the case of sine wave voltage pulse injection where, during a period $T$, $V(t)=\sum_{k} b_{k}(h/eT)(1-\cos2\pi(t-kT)/T))$. One finds $\Delta N_{exc}/\langle q\rangle=0.05476$ (and $\langle q\rangle=e/2)$ while for the periodic case one has $\Delta N_{exc}/\langle q\rangle=0.028$ (and $<q>=1$).  This confirms that sine wave pulses produce less extra excitations that square wave pulses but in both cases the pseudo-random binary injection generates more excitations per charge injected than the periodic injection. From this study \textit{we learn that the number of extra excitations is not only a property of a given (single) pulse shape} but depends on the way the pulses are injected. The values of the number of extra electron-hole pair excitation per pulses $\Delta N_{exc}/\langle q\rangle$ are summarized in table \ref{Tab}.
\begin{table}[htbp]
\centering
\begin{tabular}{|l|c|c|c|r|}
  \hline
  injection & $\langle q\rangle$ &Lorentzian & sine pulse & square pulse\\
  \hline
  \hline
  periodic & 1 & 0 & 0.028 & 0.1082\\
  \hline
  pseudo-random & 0.5 & 0 & 0.0548 & 0.1289\\
  \hline
\end{tabular} 
 \caption{ $\Delta N_{exc}/\langle q\rangle$ number of electron-hole pairs per mean injected charge $\langle q\rangle$. Equal bit probability is assumed here.} \label{Tab}
\end{table}

\subsection{\label{sec:level4B}B Fractional charge pulses}
 The discrepancy between periodic and binary injection becomes more pronounced if we consider non-integer pulses. Indeed, the injection of pulses carrying non-integer charge can never give rise to a minimal excitation states, even in the case of Lorentzian voltage pulses. For non minimal excitation pulses, like sine or square wave pulses, this also manifests as a cusp in the electron-hole number per pulse versus pulse charge when the charge crosses an integer value, see Figure 6 in \cite{Dubo12}.  This strong result, called Dynamical Orthogonality Catastrophe (DOC) in  \cite{coherentstates} tells us that injecting a non-integer charge in a non-interacting Fermi system can be done only at the expense of a large superposition of electron and hole excitations. 
 
 In particular, the number of excited particle-hole pairs detected
over a large time interval $t$ diverges as $\ln (t/W)$ \cite{coherentstates,Sherkunov2010,Knap2012}, where W is the time for switching the perturbation and is typically the width of Lorentzian pulses here. In general, more space between pulses gives more freedom to the excited Fermi sea to create extra excitations. Thus, for binary injection, the average waiting time between injected wave-packets being larger than in the periodic case, one expects to observe more particle-hole excitation per pulses. 

Results are given here for sine wave pulses in
 Figure \ref{cata} which displays the evolution of $\Delta N_{exc}$ versus the charge per pulse for periodic and non-periodic binary injection. For non-integer charge value, the number of electron-hole excitations clearly rises signaling the Dynamical Orthogonality Catastrophe.  As a perspective, a similar study could shed light on the curious properties of fractional charge pulses \cite{Gaury14,Hofe14} and on the recently considered half-levitons \cite{MoskaPRL16} which are singular zero-energy fractional excitations minimizing noise in superconducting normal junctions \cite{Belzig16}.
\begin{figure}
  \includegraphics[width=8cm,keepaspectratio,clip]{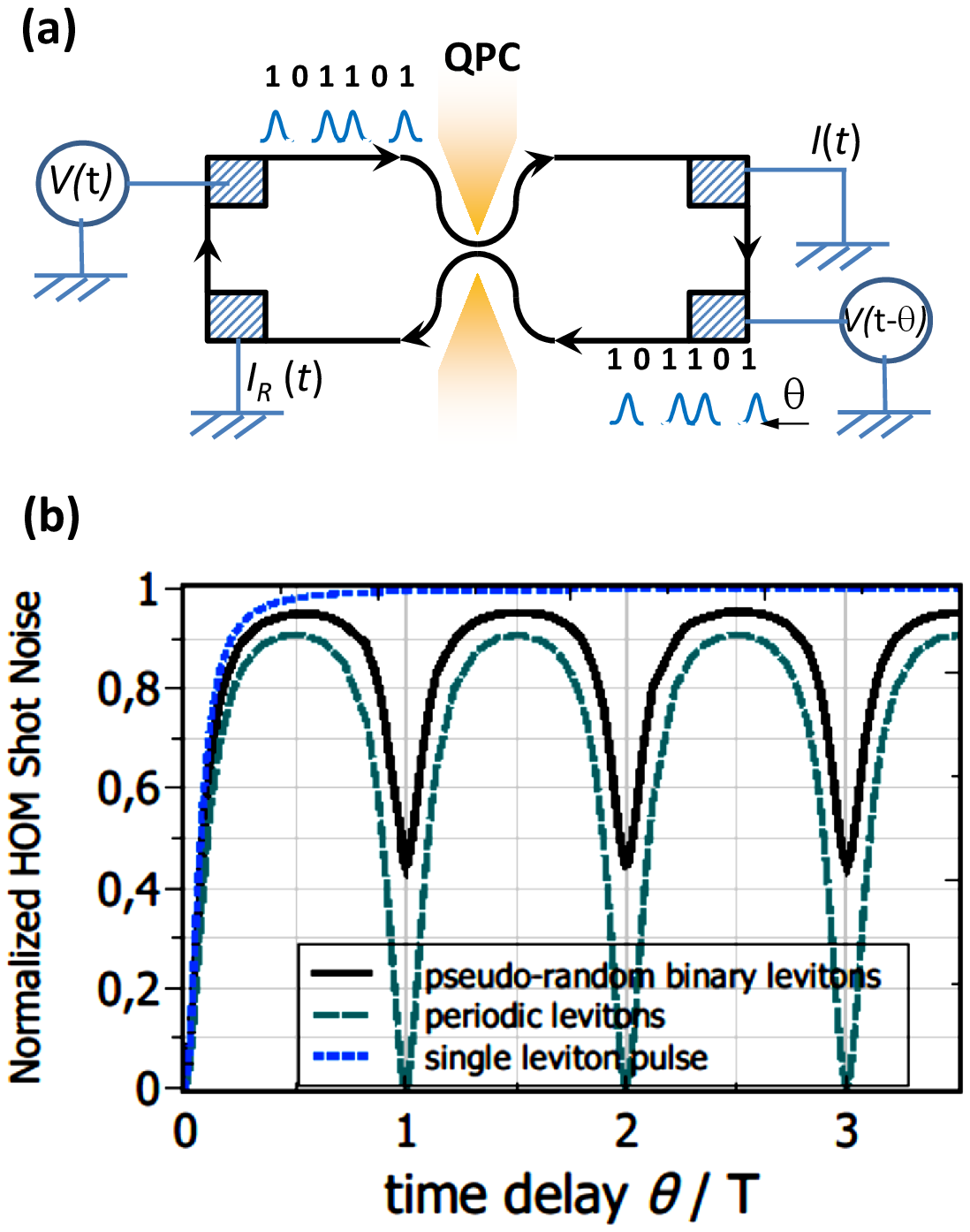}
  \caption{Hong Ou Mandel shot noise for levitons colliding in a beam-splitter with time delay $\theta$, as schematically shown in (a). In (b) The binary (solid line), periodic (dashed line) and single pulse (dotted line) injection are compared for $W=0.35 T$. While periodic injection limits the information in the range $[0,T/2]$, information for a larger time-scale is given by the binary injection whose variation at small $\theta$ are closer to that of a single pulse. Periodic HOM dips appearing for $\theta$ multiples of $T$ result from bits (0,0) and (1,1) occurrence, each with 25$\%$ probability giving a $\simeq 50\%$ HOM dip.}
  \label{HOMzero}
\end{figure}

\section{\label{sec:level5}V. Hong Ou Mandel interference with pseudo-random binary pulses}
Finally we address electronic Hong Ou Mandel (HOM) correlations where identical binary sequences of Lorentzian pulses are applied on opposite contacts
 of a QPC forming a beam-splitter of transmission $D$, see Fig. \ref{HOMzero}(a). We introduce a time delay $\theta$ between the two voltages $V_{L}(t)=V^{bin}(t-\theta/2)$ and $V_{R}(t)=V^{bin}(t+\theta/2)$. The measure of the HOM interference is given by the noise $S_{I}^{HOM}(\theta) \propto (1- |\langle\psi(\theta)|\psi(0)\rangle|^{2})$ observed in the current fluctuation of the output leads resulting from two-electron partitioning, as shown for periodic electron injection in \cite{Dubo13,Bocq13}. The time correlation function $\langle \overline{C(\tau, \theta )}\rangle$ enabling calculation of $P(\varepsilon )$, as done in Eq.(\ref{Pesprandom}), is:
\begin{equation}\label{CtauTheta}
   \langle \overline{C(\tau, \theta )}\rangle =1+I(\tau ,\theta )+ I(\tau ,-\theta )^{\ast}
\end{equation}
\begin{equation}
    \begin{split}
    I(\tau, \theta) &= \frac{i}{2}(\cos2\pi(\tau+\theta/2-iw)-\cos2\pi t^{+})\times ...\\
    &... \frac{(\cos2\pi(\tau+\theta/2+iw)-\cos2\pi t^{-})}{\sin2\pi(\tau-\theta/2+iw)\sin2\pi\tau \sin2\pi(\theta/2-iw)}
    \end{split}
\end{equation}
where:
\begin{equation}
    t^{\pm} = \bigg( \tau^{2}+(\theta/2)^{2}-w^{2} \pm \sqrt{4\tau^{2}((\theta/2)^{2}-w^{2}) - (w\theta/2)^{2}}\bigg) ^{1/2}
\end{equation}
The HOM noise is shown in Figure \ref{HOMzero}. For $\theta=0$ electrons emitted by identical sequences are undistinguishable at all time. The Fermi statistics leads to perfect antibunching and 100$\%$ noise suppression is found. We see replica of noise suppression for $\theta=kT$, 50$\%$ deep, which correspond to (0,0) and (1,1) bit events each occurring with 25$\%$ probability. These dips could be reduce by one half using binary Barker codes \cite{Barker53} characterized by a sharply peaked correlation function $\langle b_{k}b_{k+p} \rangle $ at $p=0$.  The HOM noise of periodic levitons is shown for comparison as well as the single pulse HOM noise. The pseudo-random binary injection is in between and provides information on the leviton not limited to $\theta\leq T/2$.
\begin{figure}
  \includegraphics[width=8cm,keepaspectratio,clip]{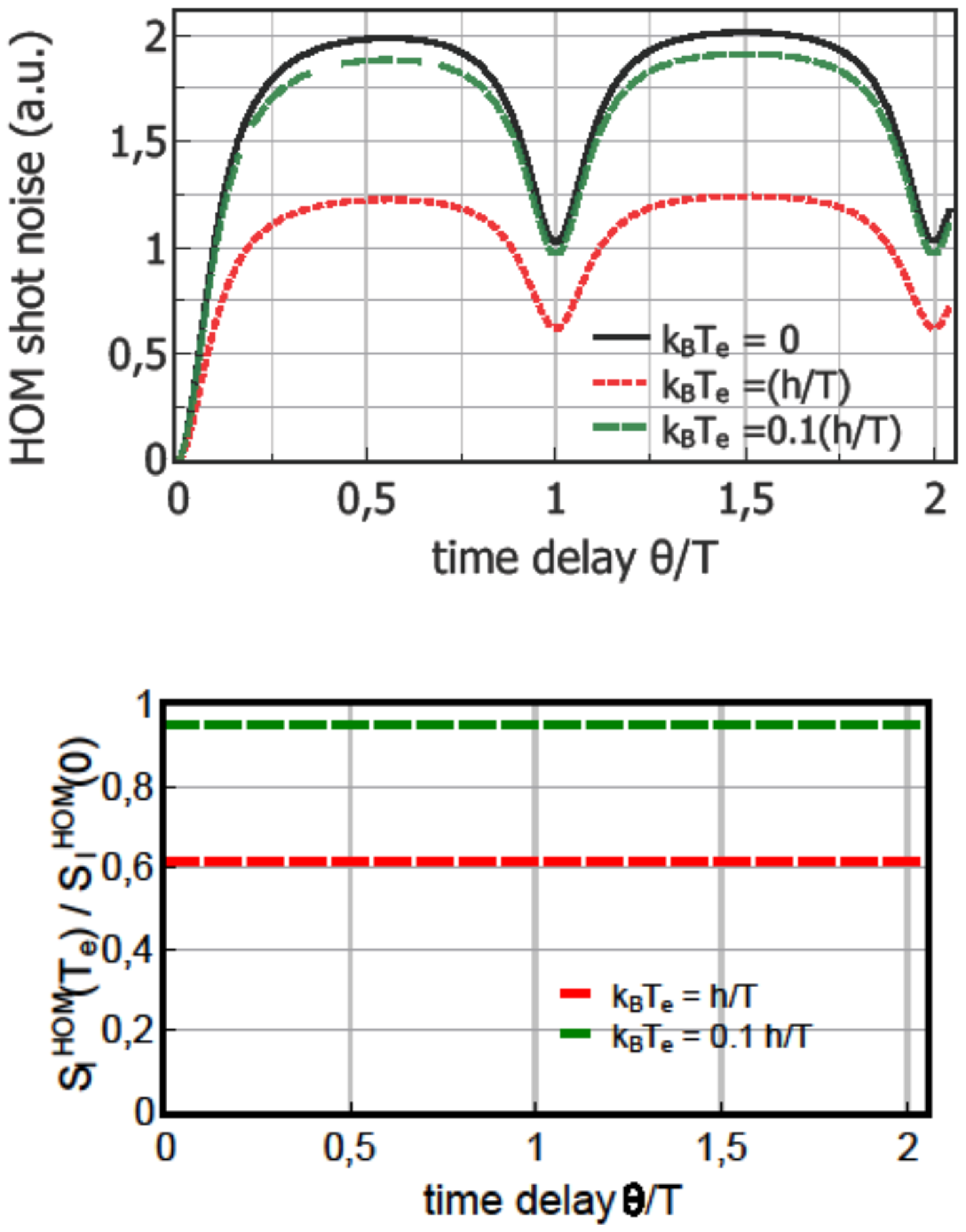}
  \caption{Upper graph: Hong Ou Mandel shot noise for levitons colliding in a beam-splitter with time delay $\theta$ and $W=0.05T$ for different electronic temperatures $T_{e}$.
  Lower graph : The ratio between curves show that they are homothetic: remarkably the shape of the HOM noise curve versus time delay is not affected by temperature, except an overall reduction factor, as already observed for periodic injection.}
  \label{HOMtempe}
\end{figure}
To complete this study we generalize to random injection the remarkable result, found for periodic injection in \cite{Dubo12} and observed in \cite{Glattli16}, that the HOM leviton noise shape versus $\theta$ is not affected by temperature, a property not shared by sine or square wave charge pulses. This is a robust property which has also been theoretically observed in \cite{Rech17} in the interacting regime of the Fractional Quantum Hall Effect where chiral edge channels form  Luttinger liquids. This confirms the prediction of \cite{Moska16} that finite temperature, unlike de-coherence, does not affect leviton HOM correlations. This is shown in Figure \ref{HOMtempe} which demonstrates the homothetic property of HOM noise curves for different temperature.
Finally, it will be interesting to consider the HOM noise resulting from interference of pseudo random multiple charge levitons, that it $b_{k}=0$, 1, 2, 3. While we expect zero HOM noise at zero delay, the HOM dip replica should be different from the case of single charge consider here. In the periodic case, the HOM noise of doubly charge levitons interfering in a Quantum Point Contact has been theoretically derived and experimentally observed in \cite{Glattli17b}. It would be interesting to look at the effect of pseudo-rando injection for  multiple charge leviton HOM interference in the Fractional Quantum Hall regime in view of the recent theoretical observation of crystallized leviton wavepackets in this interacting regime \cite{Rone17}. 

To conclude we have generalized the Floquet scattering approach to the non-periodic driving regime of a quantum conductor. We have given the analytic form of the time correlation function for the binary pseudo-random injection of levitons and for other charge pulses carrying integer charge. We have found the counterintuitive result that a random, but regular, injection does not spoils the coherence of wavepackets as signaled by  finite off-diagonal elements of the energy density matrix. For non-integer charge pulses we found that the number of neutral electron-hole excitations accompanying the pulses increases compared with the periodic injection as a result of the Dynamical Orthogonality Catastrophe. Considering the Hong Ou Mandel interference of pseudo-random integer levitons we found additional HOM dips resulting from the statistical antibunching of electrons and we have confirmed the striking  robust temperature independent HOM shape versus time delay. 

Considering random injection provides a new tool leading to tractable theoretical results to give new information not accessible in periodic injection. Further studies may include varying the bit injection probability to provide a systematic study of the DOC or use the binary injection to simulate flying qubit operation. We think this work may be also of interest for the rising community studying Floquet driven Hamiltonian and give them new perspectives. We hope that this work will inspire new studies exploiting spread spectrum approaches to get new information on electronic or non-electronic quantum systems.

The ANR FullyQuantum and ERC Proof of Concept C-Levitonics grants, and useful discussions with J. Palicot, Y. Louet, C. Moy, H. Fares, and C. Bader, and with I. Safi and D. Ferraro and with Nanoelectronics members are acknowledged.

\bibliography{References}

\begin{references}



\bibitem{Ji03} Y. Ji, Y. Chung, D. Sprinzak, M. Heiblum, D. Mahalu, and H.
Shtrikman, \textit{An electronic Mach-Zehnder interferometer }, Nature London \textbf{422}, 415 (2003)

\bibitem{Roulleau08} P. Roulleau, F. Portier, P. Roche, A. Cavanna, G. Faini, U.
Gennser, and D. Mailly, \textit{Direct Measurement of the Coherence Length of Edge States in the Integer Quantum Hall Regime},  Phys. Rev. Lett. \textbf{100}, 126802 (2008)

\bibitem{flying_quBits1} A. Bertoni, P. Bordone, R. Brunetti, C. Jacoboni, and
S. Reggiani, \textit{Quantum logic gates based on coherent electron transport in quantum wires} \emph{Phys. Rev. Lett.} \textbf{84}, 5912-5915 (2000).

\bibitem{flying_quBits2}  R.
Ionicioiu, G. Amaratunga, and F. Udrea, \textit{Quantum Computation with ballistic electrons} \emph{Int. J. Mod. Phys.}
\textbf{15}, 125-133 (2001).

\bibitem{Herm11}%
S. Hermelin, S. Takada, M. Yamamoto, S. Tarucha, A. D. Wieck, L. Saminadayar, C. B\"{a}uerle, and T. Meunier, \textit{Electrons surfing on a sound wave as a platform for quantum optics with flying electrons,}
Nature 477 435 (2011)

\bibitem{McNe11}%
R. McNeil, M. Kataoka, C. Ford, C. Barnes, D. Anderson, G. Jones, I. Farrer, and D. Ritchie, \textit{On-demand single-electron transfer between distant quantum dots,} Nature 477, 439 ? 442 (2011)

\bibitem{Flet13}%
J. D. Fletcher, P. See, H. Howe, M. Pepper, S. P. Giblin, J. P. Griffiths, G. A. C. Jones, I. Farrer, D. A. Ritchie, T. J. B. M. Janssen, and M. Kataoka, \textit{Clock-Controlled Emission of Single-Electron Wave Packets in a Solid-State Circuit,} 
Phys. Rev. Lett. 111 216807 (2013)

\bibitem{Ryu16}%
S. Ryu, M. Kataoka, and H.-S. Sim,\textit{ Ultrafast Emission and Detection of a Single-Electron Gaussian Wave Packet: A Theoretical Study,} 
Physical Review Letters, 117, 146802 (2016)

\bibitem{Ubbe14}%
Niels Ubbelohde, Frank Hohls, Vyacheslavs Kashcheyevs, Timo Wagner, Lukas Fricke, Bernd K\"{a}stner, Klaus Pierz, Hans W. Schumacher, and Rolf J. Haug, \textit{Partitioning of on-demand electron pairs,}
Nat. Nanotechnol. 10, 46 (2015).

\bibitem{Feve07}%
G. F\`{e}ve, A. Mahe, J-M Berroir, T Kontos, B. Placais, D C Glattli, A Cavanna, B
Etienne, and Y Jin, An on-demand coherent single-electron source, Science 316 , 1169 - 1172 (2007)

\bibitem{Dubo13}%
J. Dubois, T. Jullien, F. Portier, P. Roche, A. Cavanna, Y. Jin,
W. Wegscheider, P. Roulleau, D. C. Glattli, \textit{Minimal-excitation states for electron quantum optics using levitons,} Nature. \textbf{502} (2013) 659 - 663

\bibitem{Jull14}%
T. Jullien, P. Roulleau, B. Roche, A. Cavanna, Y. Jin, D.C. Glattli, \textit{Quantum tomography of an electron,}  Nature. \textbf{514} (2014)
603 - 607.

\bibitem{coherentstates} L. S. Levitov, H. Lee, and G. Lesovik,\textit{Electron counting statistics and coherent states of electric current,} J. Math. Phys. \textbf{37}, 4845 (1996);

\bibitem{Ivan97}%
 D. A. Ivanov, H. W. Lee, and L. S. Levitov, \textit{Coherent states of alternating current,}
Phys. Rev B \textbf{56},  (1997) 6839.

\bibitem{lebePRB05} Lebedev, A. V., Lesovik, G. V. and Blatter, G., \textit{Generating spin-entangled electron pairs in normal conductors using voltage pulses,}  Phys. Rev. B \textbf{72}, (2005) 245314.

\bibitem{Keeling06} J. Keeling, I. Klich, and L. S. Levitov, \textit{Minimal excitation states of electrons in one-dimensional wires,} Phys. Rev. Lett. \textbf{97}, (2006) 116403.

\bibitem{Hassler08} F. Hassler, M. V. Suslov, G. M. Graf, M. V. Lebedev, G. B. Lesovik, and G. Blatter, \textit{Wave-packet formalism of full counting statistics,} Phys. Rev. B \textbf{78}, (2008) 165330.

\bibitem{Janine09} Janine Splettstoesser, Michael Moskalets, and Markus B\"{u}ttiker, \textit{Two-particle nonlocal Aharonov-Bohm effect from two single-particle emitters,} 
Phys. Rev. Lett. \textbf{103}, 076804 (2009)

\bibitem{Moskalets11} Michael Moskalets and Markus B\"{u}ttiker, \textit{Spectroscopy of electron flows with single and two-particle emitters,} 
Phys. Rev. B \textbf{83}, 035316 (2011)

\bibitem{Sherkunov2010} Y. B. Sherkunov, A. Pratap, B. Muzykantskii, and N. d’Ambrumenil,\textit{ Quantum point contacts, full counting statistics and the geometry of planes,} 
Opt. Spectrosc. 108 466 (2010)

\bibitem{Knap2012} Michael Knap, Aditya Shashi, Yusuke Nishida, Adilet Imambekov, Dmitry A. Abanin, and Eugene Demler, \textit{Time-Dependent Impurity in Ultracold Fermions: Orthogonality Catastrophe and Beyond,} Phys. Rev. X 2, 041020  (2012)


\bibitem{Parm12}%
F. D. Parmentier, E. Bocquillon, J.-M. Berroir, D. C. Glattli, B. Plaçais, G. F\`{e}ve, M. Albert, C. Flindt, and M. B\"{u}ttiker,\textit{ Current noise spectrum of a single-particle emitter: Theory and experiment,} Phys. Rev. B 85, 165438 (2012)

\bibitem{Bocq13}%
    E. Bocquillon, V. Freulon, J.-M Berroir, P. Degiovanni, B. Plaçais, A. Cavanna, Y. Jin, G. F\`{e}ve, \textit{Coherence and Indistinguishability of Single Electrons Emitted by Independent Sources,} Science  339, 1054 - 1057 (2013)

\bibitem{Moska15} M. Moskalets, \textit{First-order correlation function of a stream of single-electron wave packets,}  Phys. Rev. B \textbf{91}, 195431 (2015)
\bibitem{Gren11} C. Grenier, R. Herv\'{e}, G. F\`{e}ve, P. Degiovanni, \textit{Electron quantum
optics in quantum hall edge channels}, Mod. Phys. Lett. B. \textbf{25} 1053–1073 (2011).

\bibitem{Haack13} G. Haack, M. Moskalets, M. B\"{u}ttiker, \textit{Glauber coherence of single-electron sources}, Physical Review B. \textbf{87} 201302 (2013)

\bibitem{Gaury14}
B. Gaury and X. Waintal , \textit{Dynamical control of interference using voltage pulses in the quantum regime , } Nature Communications, 5, 3844 (2014)


\bibitem{Dubo12}%
J. Dubois, T. Jullien, C. Grenier, P. Degiovanni, P. Roulleau, and D. C. Glattli, \textit{Integer and fractional charge Lorentzian voltage pulses analyzed in the framework of photon-assisted shot noise.},  Phys. Rev. B \textbf{88}, 085301 (2013)

\bibitem{MartPRB92} Th. Martin and R. Landauer, \textit{Wave-packet approach to noise in multichannel mesoscopic systems,} Phys. Rev. B \textbf{45}, (1992) 1742.

\bibitem{ButtPRB92} M. B\"{u}ttiker, \textit{Scattering theory of current and intensity noise correlations in conductors and wave guides,} Phys. Rev. B \textbf{46}, (1992)  12485.

\bibitem{Blant00} Ya. M. Blanter and M. B\"{u}ttiker, \textit{Shot noise in mesoscopic conductors,} Phys. Rep. \textbf{336}, 1 (2000)

\bibitem{Grenier11} C. Grenier, R. Herv\'{e}, E. Bocquillon, F. D. Parmentier,
B. Pla\c{c}ais, J. M. Berroir, G. F\`{e}ve, and P. Degiovanni, \textit{Single-electron quantum tomography in quantum Hall edge channels,}  New Journal of Physics \textbf{13}, 093007 (2011).

\bibitem{Ferra13} D. Ferraro, A. Feller, A. Ghibaudo, E. Thibierge, E. Bocquillon, G. Fève, Ch. Grenier, and P. Degiovanni, \textit{Wigner function approach to single electron coherence in quantum Hall edge channels} Phys. Rev. B \textbf{88}, 205303 (2013)

\bibitem{Safi14} I. Safi, \textit{Time-dependent Transport in arbitrary extended driven tunnel junction} 
 arXiv:1401.5950 [cond-mat.mes-hall] (2014)

\bibitem{Moska02}
M. Moskalets and M. B\"{u}ttiker, \textit{Floquet scattering theory of quantum pumps,}
Phys. Rev. B 66, 205320 (2002) 

\bibitem{Dahlhaus11} . P. Dahlhaus, J. M. Edge, J. Tworzydlo, and C. W. J. Beenakker, \textit{Quantum Hall effect in a one-dimensional dynamical system, } Phys. Rev. B 84, 115133 (2011)

\bibitem{MartinHalperin17} I. Martin, G. Refael and B. Halperin,\textit{ Topological Frequency Conversion in Strongly Driven Quantum Systems}, Phys. Rev. X 7, 041008 (2017)

\bibitem{Lesovik94} G. B. Lesovik and L. S. Levitov,\textit{ Noise in an ac biased junction: Nonstationary Aharonov-Bohm effect,}  Phys. Rev. Lett. \textbf{72}, 538 (1994).

\bibitem{Pedersen98} M. H. Pedersen and M. B\"{u}ttiker, \textit{Scattering theory of photon-assisted electron transport}, Phys. Rev. B \textbf{58} (1998) 12993.

\bibitem{Rychkov05} V. S. Rychkov, M. L. Polianski, and M. B\"{u}ttiker,\textit{ Photon-assisted electron-hole shot noise in multiterminal conductors,}  Phys. Rev. B \textbf{72}, 155326 (2005)

\bibitem{Vanevic07} M. Vanevic, Y. V. Nazarov, and W. Belzig, \textit{Elementary events of electron transfer in a voltage-driven quantum point contact,}  Phys. Rev. Lett. \textbf{99}, 076601 (2007)

\bibitem{Bagrets07} D. Bagrets and F. Pistolesi, \textit{Frequency dispersion of photon-assisted shot noise in mesoscopic conductors,}  Phys. Rev. B \textbf{75}, 165315 (2007)

\bibitem{Proakis} J.G. Proakis and M. Salehi, Digital Communications, 5th edition, New York: McGraw-Hill, 2008.

\bibitem{Fares17} H. Fares, D. C. Glattli, Y. Louet, J. Palicot, P. Roulleau and C. Moy, \textit{Power Spectrum density of Single Side band CPM using Lorentzian frequency pulses,} IEEE Wireless Communication Letters, Volume: 6, Issue: 6, p 786-789 (2017) ;DOI: 10.1109/LWC.2017.2742505.

\bibitem{MoskaPRL16} Michael Moskalets,\textit{ Fractionally Charged Zero-Energy Single-Particle Excitations in a Driven Fermi Sea,}  Phys.  Rev. Lett. \textbf{117}, 046801 (2016)

\bibitem{Hofe14}
P. P. Hofer and C. Flindt,\textit{ Mach-Zehnder interferometry with periodic voltage pulses, } Phys. Rev. B \textbf{90}, 235416 (2014) 

\bibitem{Belzig16}  Wolfgang Belzig, Mihajlo Vanevic, 
\textit{Elementary Andreev processes in a driven superconductor–normal metal contact}, 
Physica E: Low-dimensional Systems and Nanostructures, 75, p22-27 (2016)

\bibitem{Barker53}  Barker, R. H. (1953). \textit{Group Synchronizing of Binary Digital Systems}, Communication Theory. London: Butterworth. pp. 273–287

\bibitem{Glattli16} D.C. Glattli and P. Roulleau, \textit{Hanbury-Brown Twiss noise correlation with time
controlled quasi-particles in ballistic quantum conductors,}  Physica E: Low-dimensional Systems
and Nanostructures, 82, 99-105 (2016)

\bibitem{Moska16} Michael Moskalets, G\'{e}raldine Haack, \textit{Single-electron coherence: Finite temperature versus pure dephasing, } special issue in Physica E on "Frontiers in quantum electronic transport" - in memory of Markus B\"uttiker,  Physica E \textbf{75}, 358-369 (2016) 

\bibitem{Rech17} J. Rech, D. Ferraro, T. Jonckheere, L. Vannucci, M. Sassetti, T. Martin, \textit{Minimal excitations in the fractional quantum Hall regime }, Phys. Rev. Lett. \textbf{118}, 076801 (2017)

\bibitem{Glattli17b} D.C. Glattli and P. Roulleau, \textit{Levitons for electron quantum optics}, 
Phys. Status Solidi B 254, No. 3, 1600650 (2017) /DOI 10.1002/pssb.201600650

\bibitem{Rone17} Flavio Ronetti, Luca Vannucci, Dario Ferraro, Thibaut Jonckheere, Jérôme Rech, Thierry Martin, Maura Sassetti, \textit{Crystallization of Levitons in the fractional quantum Hall regime},  	arXiv:1712.07094 [cond-mat.mes-hall] (2017)













\end{references}

\end{document}